\documentclass[prc,amsmath,twocolumn,superscriptaddress,showpacs,showkeys]{revtex4} 
\usepackage{graphicx}
\usepackage{amssymb} 
\usepackage{amsmath} 

\usepackage[usenames,dvips]{color}

\renewcommand*{\eqref}[1]{Eq.~(\ref{eq:#1})}

\newcommand*{\eqlab}[1]{\label{eq:#1}}
\newcommand*{\figref}[1]{Fig.~(\ref{fig:#1})}
\newcommand*{\figlab}[1]{\label{fig:#1}}

\newcommand*{\secref}[1]{Section~\ref{sec:#1}}
\newcommand*{\seclab}[1]{\label{sec:#1}}

\def\VYP#1#2#3{{\bf #1}, #3 (#2)}  

\def\PLB#1#2#3{Phys.~Lett.~B~\VYP{#1}{#2}{#3}}

\def\PRD#1#2#3{Phys.~Rev.~D~\VYP{#1}{#2}{#3}}
\def\PRE#1#2#3{Phys.~Rev.~E~\VYP{#1}{#2}{#3}}
\def\PRL#1#2#3{Phys.~Rev.~Lett.~\VYP{#1}{#2}{#3}}

\newcommand{\etal}{\mbox{\textit et al.}}                       %

\newcommand{\Omit}[1]{}

\begin{document}

\title{A new limit on the Ultra-High-Energy Cosmic-Ray flux with the Westerbork Synthesis Radio Telescope}

\author{S. ter Veen} \affiliation{ Department of Astrophysics, IMAPP, Radboud University, 6500 GL Nijmegen, The Netherlands}
\email{s.terveen@astro.ru.nl}
\thanks{Based on the Master thesis of S.~ter Veen, University of Groningen, 2008, unpublished}
\author{S. Buitink} \affiliation{Lawrence Berkeley National Laboratory, Berkeley, California 94720, USA}
\author{H. Falcke} \affiliation{ Department of Astrophysics, IMAPP, Radboud University, 6500 GL Nijmegen, The Netherlands} \affiliation{ASTRON, Dwingeloo, P.O. Box 2, 7990AA Dwingeloo, The Netherlands}
\author{C.W. James} \affiliation{ Department of Astrophysics, IMAPP, Radboud University, 6500 GL Nijmegen, The Netherlands}
\author{M. Mevius}  \affiliation{Kernfysisch Versneller Instituut, University of Groningen, 9747 AA, Groningen, The Netherlands}
\author{O. Scholten}  \affiliation{Kernfysisch Versneller Instituut, University of Groningen, 9747 AA, Groningen, The Netherlands}
\author{K. Singh} \affiliation{Kernfysisch Versneller Instituut, University of Groningen, 9747 AA, Groningen, The Netherlands}
\affiliation{Vrije Universiteit Brussel, Dienst ELEM, B-1050 Brussels, Belgium}
\author{B. Stappers}  \affiliation{Jodrell Bank Centre for Astrophysics, School of Physics and Astronomy, The University of Manchester, Manchester M13 9PL, UK }
\author{K.D. de~Vries} \affiliation{Kernfysisch Versneller Instituut, University of Groningen, 9747 AA, Groningen, The Netherlands}


\date{\today}

\begin{abstract}
A particle cascade (shower) in a dielectric, for example as initiated by an ultra-high energy cosmic ray, will have an excess of electrons which will emit coherent \v{C}erenkov radiation, known as the Askaryan effect. In this work we study the case in which such a particle shower occurs in a medium just below its surface. We show, for the first time, that the radiation transmitted through the surface is independent of the depth of the shower below the surface when observed from far away, apart from trivial absorption effects. As a direct application we use the recent results of the NuMoon project, where a limit on the neutrino flux for energies above $10^{22}$\,eV was set using the Westerbork Synthesis Radio Telescope by measuring pulsed radio emission from the Moon, to set a limit on the flux of ultra-high-energy cosmic rays.
\end{abstract}

    \maketitle

\section{Introduction}

Cosmic rays, charged particles traveling through the universe, have been observed at energies ranging from GeV to above $10^{20}$\,eV~\cite{Abraham:2008ru}. The highest observed energies lie well above energies that can be reached by particle accelerators on Earth. These particles are of particular interest to astrophysics and particle physics to answer fundamental questions about cosmic acceleration mechanisms and particle interactions.
The Astrophysics interest stems from the fact that ultra-high-energy (UHE) cosmic rays, particles with energies above $6 \cdot 10^{19}$~eV have two important properties.
They are less deflected by the galactic magnetic fields and therefore contain more information about their source of origin~\cite{PAO-Correlation}, which is still unknown. It should be noted that the real energy threshold at which this becomes apparent is dependent on the type of particle and scales with their rigidity.
In addition their sources are not too far from Earth since they interact with the cosmic microwave background to produce pions and therefore experience substantial energy loss over distances of the order of 50\,Mpc. This is known as the GZK-effect~\cite{Greisen}. To detect these particles requires a large collecting area, because the flux at 60\,EeV is only $1 / \rm{km}^2 / \rm{century}$ and drops off faster than $\rm{E}^{-2.6}$. This leads to detectors like the Pierre Auger Observatory with a total collecting area of 3000\,km$^2$~\cite{PAO}. However, to detect particles above $10^{21}$\,eV a thousand-fold increase in collecting area is expected to be needed.

As first proposed in Ref.~\cite{dz89} the Moon is a suitable candidate with an area of $10^7\,\rm{km}^2$. Detection is based on the fact that when a high-energy particle interacts an avalanche reaction occurs creating a cascade (shower) of particles. The number of particles near the maximum is roughly proportional to the total energy in the avalanche and is of the order of $10^{12}$ for
$10^{21}$\,eV~\cite{Alvarez-MunizZas98}. In this particle shower there will be a net
excess of electrons due to the knock-out of atomic electrons by shower positrons and high energy photons (Compton scattering). Simulations show that this excess amounts to about 20\%~\cite{ZasHalzenStanev92}, consistent with experimental observations~\cite{SaltzbergGorham01}.
Since all particles move with almost the light velocity they are closely bunched in the longitudinal as well as the lateral direction. In a
dielectric this will result in the emission of coherent \v{C}erenkov radiation at
wavelengths that are larger than the typical dimensions of the charge cloud; this is known as the Askaryan effect \cite{Askaryan}. For materials
like ice, salt or lunar regolith this implies coherent radiation at frequencies of 3\,GHz
and less~\cite{ZasHalzenStanev92} and this mechanism is used in several experiments to detect
high-energy cosmic neutrinos. Well known examples are the  ANITA~\cite{anita}, GLUE~\cite{glue}, LUNASKA~\cite{lunaska},
and NuMoon~\cite{Sch09} experiments.

For a particle shower deep inside the dielectric, such as is usually the case
for neutrino-induced showers, the picture for the Askaryan effect clearly applies.
The subject of the present work is to investigate the emitted radiation for the case that
the particle shower occurs very close to the boundary between the dielectric and vacuum. In
this case one can imagine that a dielectric layer of minimal thickness between the shower
and vacuum is required for \v{C}erenkov emission to occur. This relates to the concept of a
formation zone/formation length for \v{C}erenkov radiation as was introduced in Ref.~\cite{Tak94} and used in
calculations of the acceptance for detecting high-energy cosmic particles~\cite{Gor01}.

Since neutrinos are weakly interacting particles they can traverse many hundreds of kilometers in dense materials before interacting. For neutrino-induced showers the issue of a possible formation length
is thus not essential since they typically interact deep in the Moon, compared to which
any formation length would be negligible. Cosmic rays are particles interacting via the strong interaction and thus induce a reaction well within a meter from the surface of a dense material.
For cosmic-ray induced showers the possible existence of a formation length for the
emission of electromagnetic radiation could make the difference between being able to observe
the shower or not especially at wave lengths of the order of a few meters.
The study we present in this paper indicates that for the motion of charged particles inside a dielectric the concept of a formation zone does not apply. We will show that due to the finiteness of a particle trajectory inside a dense medium the
radiation detected in vacuum is independent of the depth of the trajectory below
the surface, other than for absorption in the medium.
This situation is very different from that for the opposite geometry, an effectively infinite electron beam
in vacuum inducing \v{C}erenkov emission in a dielectric, as has
been investigated theoretically in Ref.~\cite{Ulrich6667} and confirmed in
experiments~\cite{Tak00}. The essence of the difference lies in the fact that for particle showers in a medium the track length is necessarily finite for which case \v{C}erenkov and bremsstrahlung emission cannot really be distinguished, as was already noted in Refs.~\cite{Law65,Afa99}.

Using the new finding of an absence of a formation length for radiation emitted by a shower
in a dense material we have calculated the detection probability for observing radio
emission from cosmic-ray impacts on the lunar surface. The idea to observe this type of
emission from the Moon with radio telescopes was first proposed by~\cite{dz89} and the
first experimental endeavors in this direction were carried out with the Parkes telescope~\cite{parkes96}. It was shown in Ref.~\cite{Sch06} that observations
in the frequency range of 100-200\,MHz (as suggested by Ref.~\cite{FalckeGorham}) maximizes the detection probability with small
loss of sensitivity. This was used in recent lunar observations with the Westerbork Synthesis Radio Telescope (WSRT) to set a new limit on
the flux of ultra-high energy (UHE) neutrinos~\cite{Sch09}. Combining these observational data
and our present calculation of the detection efficiency we set a new limit on the flux of
cosmic rays at energies in excess of $10^{22}$\,eV. Such a limit is of interest since
recent results of the PAO indicate a steepening in the cosmic ray spectrum at the
Greisen-Zatsepin-Kuzmin (GZK) energy of $6\cdot 10^{19}$~eV~\cite{PAO10}. The flux of cosmic
rays above this energy will be a clear indication of nearby sources and the nature thereof. In future observations with new generation, more sensitive, synthesis telescopes this method may be used to measure the flux at GZK energies.

\section{Shallow Showers}

The argument that a formation-zone effect will reduce the emission for shallow showers
hinges on two intuitive points: one, that as the distance from the cascade to the surface decreases, the emission increasingly becomes insensitive to the presence of the dielectric; and two, that the \v{C}erenkov radiation from
such a cascade in vacuum (a gedankenexperiment) is zero. In this section, we proceed to disprove the second of these two points, and establish that current calculational methods, as used in programs to estimate the radiated intensity from particle showers in
the Moon, correctly model shallow showers. As discussed at the end of this section, this is intuitively understandable, because for a finite track the actual radiation is mostly due to acceleration and deceleration of the net charge at the end-points~\cite{Law65,Afa99}. In this process coherent radiation is emitted, known as the Askaryan effect~\cite{Askaryan}. As is well known~\cite{Law65,Afa99} for finite trajectories one cannot distinguish between \v{C}erenkov radiation and Bremsstrahlung.

We will first (in \secref{far}) outline the standard approach used in the calculation of
the emitted \v{C}erenkov-radiation intensity for neutrino induced showers, and in \secref{dff}
describe how this has been used to calculate
the acceptance for the GLUE~\cite{glue}, LUNASKA~\cite{lunaska},
NuMoon~\cite{Sch09}, and other experiments. In \secref{zero} we show that
this double far-field approach (shower far below the surface and the observer far away from the surface) predicts that a deep cascade (excluding absorption) will produce the same observable radiation as a shower developing in the vacuum above the lunar surface.
Since a near-surface cascade (i.e.\ a cosmic ray induced shower) is intermediary to these two extremes one thus expects the same radiation as from a deep cascade barring absorbtion effects.
To show this at a more rigorous level we present in \secref{exact} an exact treatment which verifies
that the current double far-field approximation gives the correct results even for shallow showers.

\subsection{\v{C}erenkov radiation from finite particle tracks} \seclab{far}

The calculation of the \v{C}erenkov radiation arising from finite particle tracks is usually
performed assuming the observer is in the far-field in an infinite dielectric medium.
According to classic electromagnetic theory a
shower with a current density of $J_z(\mathbf{r},t)= c q(z) \delta(\mathbf{r}-c t\hat{z})$ and
$J_x=J_y=0$ leads in the far field (distance much greater than the shower dimension) to a
vector potential
\begin{equation}
A_z(\mathbf{R},\omega)=\frac{e^{ikR}}{\sqrt{2\pi}c\, R}
\int_{-\infty}^\infty q(z)\, e^{-iz \omega(n \cos \theta_s - 1)/c} dz \;, \eqlab{Arw}
\end{equation}
where $\theta_s$ is the angle between $\mathbf{R}$ and $\hat{z}$ as shown in \figref{diagram}, and $n$ is the index of refraction of the medium. The electric field can be calculated to be $E=\omega A_z \sin\theta_s$ where $\vec{E}$ lies in the plane spanned by
the shower and $\mathbf{R}$, thus
\begin{align}
E(\mathbf{R},\omega)=& \omega \sin\theta_s\, \frac{e^{i\omega n R/c}}{\sqrt{2\pi}c\, R} \times
 \nonumber \\
 &\int_{-\infty}^\infty q(z)\, e^{-iz\omega(n \cos \theta_s - 1)/c} dz \;.
\eqlab{efarfield}
\end{align}
The function $E(\mathbf{R},\omega)$ can be calculated analytically for simple
shower profiles~\cite{Tamm39,ZasHalzenStanev92,Leh04,Sch06}, or can be taken from parameterisations of detailed simulation results~\cite{Alvarez-MunizZas98}.

\subsection{Treatment for deep cascades}  \seclab{dff}

\begin{figure}
\begin{center}
\includegraphics[width=.95\linewidth]{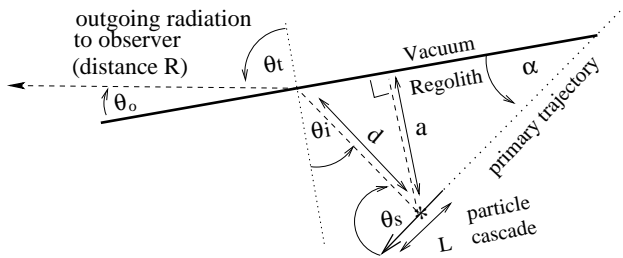}
\end{center}
\caption{Geometry of a particle cascade in the Moon. The `double-far-field' approach assumes $R \gg d \gg L$.} \figlab{diagram}
\end{figure}

Current methods to calculate the intensity of radio emission from Lunar showers
use a double-far-field approximation. Using the diagram in \figref{diagram},
this means that the shower length $L$ is taken to be small compared to the
shower depth, $a$, and the travel distance in the medium, $d$, and that the observing distance, $R$, is much larger than the shower depth. The radiation incident on the lunar surface can be taken as its far-field (i.e.\ in an infinite uniform medium)
solution. This radiation is transmitted through the surface according to simple refraction laws and the strength at the
Earth is calculated assuming the Earth-Moon distance dominates the $1/R$ term. Thus the observed radiation can be expressed as
\begin{equation}
E_{\rm obs}(R,\theta_o,\omega) = t(\theta_i, n)\, E_{\rm m}(R,\theta_s,\omega)\, e^{-d/\ell(\omega)}\;, \eqlab{method}
\end{equation}
where $t$ is the transmission coefficient, which depends on the incident angle and Lunar
refractive index, and $E_{\rm m}$, given by \eqref{efarfield}, is the infinite-medium solution of the emitted
radiation for the Moon's material properties.
$R$ is taken to be the Earth-Moon distance of approximately $3.884 \times 10^{8}$~m, and the
exponential accounts for attenuation in the medium using the frequency-dependent
attenuation-length $\ell(\omega)$.

\begin{figure}
\begin{center}
\includegraphics[width=.95\linewidth]{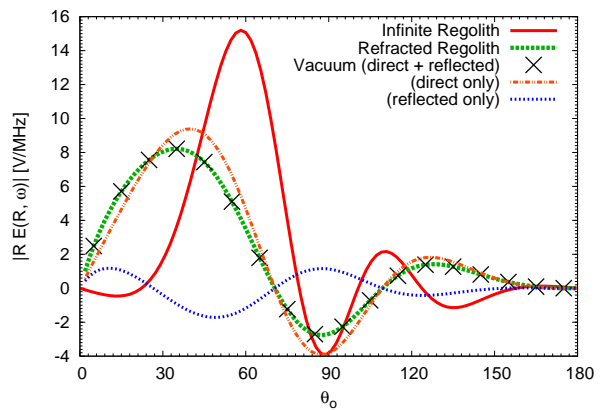}
\end{center}
\caption{[color online] Radiation pattern as function of observer angle $\theta_o$ at 150\,MHz; red drawn curve: a cascade in infinite uniform regolith; green dotted: a deep regolith cascade refracted to the vacuum; black crosses: the total observed emission from a cascade in vacuum immediately above a dielectric boundary.
The `direct' (orange) and `reflected' (blue) components of the vacuum emission are also plotted separately. The refracted
regolith and total vacuum cases are identical.} \figlab{farfield}
\end{figure}

The result of \eqref{method} is plotted in \figref{farfield} as a function of the observation
angle $\theta_o$.
The calculation is for the plane containing the shower axis and the surface normal,
so all radiation has a parallel polarisation. Also, we take the case of a cascade
parallel to the surface ($\alpha = 0$), so Snell's law becomes $n \cos \theta_s = \cos \theta_o$.
We use \eqref{efarfield} to calculate $E_{\rm m}$ for a constant charge excess $q = -10^{12}~e$ moving over a distance
$L = 3.0$~m at velocity $v=c$. The observation frequency is taken as $\nu=150$~MHz, and the regolith
refractive index as $n=1.8$. For waves diverging from a point-source, the
transmission coefficient for parallel polarisation is
\begin{eqnarray}
t_{||}(\theta_s, n) = \frac{2 \sin \theta_o}{ n \sin \theta_o + \sin \theta_s} \eqlab{tcoef} \;
\end{eqnarray}
since for $\alpha=0$ we have $\sin{\theta_s}=\cos{\theta_i}$.

\subsection{Emission in a vacuum} \seclab{zero}

The double-far-field treatment of \secref{dff} may be expected to break down as the
distance $d$ of the cascade to the surface becomes small.
To place a simple limit on near-surface effects, consider the following extreme case:
a cascade developing immediately above the surface.
In this case, the radiation seen by an observer will be that produced
in vacuum ($n=1$), but consist of both direct
and reflected components, with zero path difference due to the proximity of the
interface. For $\alpha = 0$ the radiated electric field is thus
\begin{eqnarray}
E_{\rm obs}(R,\theta_o,\omega)=\left( 1+r(\theta_o, n) \right) E_{\rm v} (\theta_o, \omega) \eqlab{vac_eqn}
\end{eqnarray}
where $E_{\rm v}$ is the radiation expected from a cascade in an infinite vacuum,
and $r$ is the Fresnel reflection coefficient (identical for plane and spherical waves).
Using \eqref{efarfield} for $E_{\rm v}$, radiation of some magnitude will be emitted.  In \figref{farfield} the result from \eqref{vac_eqn} is shown using the same parameters as used in the double-far-field approximation in \secref{dff}.
For the sake of clarity we also plot the direct and reflected contributions from \eqref{vac_eqn} separately, which add to give exactly
the same radiation pattern as that from \eqref{method} for $d = 0$. That is, the emission seen
by an observer from a cascade immediately above a dielectric boundary is exactly the same
as that from a cascade immediately below the surface, which in turn is identical to that from
a deep cascade if absorption in the medium is been ignored. It can be shown analytically in the
case $\alpha = 0$ that the two equations give identical results for an arbitrary  mix of perpendicular and parallel polarisations.

It seems counter-intuitive that any \v{C}erenkov radiation is viewed from a cascade parallel to the surface,
since radiation emitted at the \v{C}erenkov angle will be totally internally reflected.
The solution to this apparent contradiction lies in the fact that for a shower of finite length the emitted radiation has a large spread around the \v{C}erenkov angle (see for example Ref.~\cite{Law65,Sch06}). For a shower in a medium only the radiation emitted at angles larger than the \v{C}erenkov angle will penetrate the surface while for a theoretical shower in vacuum the \v{C}erenkov angle lies at zero degrees. Alternatively one may regard a shower of finite extent as corresponding to the acceleration and deceleration of charge at the beginning and end of the shower~\cite{Law65,Afa99} (equal to the appearance and disappearance of a moving charge), a picture which has also been used to explain the emission of electromagnetic radiation of showers induced in air~\cite{MGMR}.

These results indicate that there will be no change in the observed radiation as
the particle distribution induced by a UHE particle interaction lies close
to the surface.  In the following section this is shown using a more rigorous method.

\section{Exact calculation for near-surface showers}\seclab{exact}

For shallow showers, such as those from UHE cosmic-ray interactions with the lunar surface,
no far field approximation can be made for the radiation reaching the surface. For the calculation in the previous section this was an essential assumption. We avoid making this approximation by performing a complete wave-equation calculation.

In the following derivation we consider
two half-spaces divided by the plane $x=0$. For $x>0$ the refraction index is $n'$ (=1 for
vacuum). For $x<0$ the refraction index is $n$ (=1.8 for the moon). In the lower half plane
a particle with charge $Q$ and velocity $\beta$ moves from $z=-L/2$ to $z=L/2$ at $x=-a$, $y=0$, passing through $z=0$ at $t=0$ which corresponds to the same geometry as studied in the previous section.
The four-vector potential for this system is determined by the Maxwell equations. Care should be taken since the index of refraction is different for the two sides of the boundary.

In the present discussion it is sometimes easier to work in the space-time domain, sometimes with energy and momentum. The relation between the two is given by the usual Fourier transformation
\begin{equation}
\mathbf{A(r},t) = \int \!\frac{d^3 k\,d\omega}{4\pi^2} \mathbf{A(k},\omega)\, e^{i\mathbf{k\cdot r}-i\omega t} \;. \eqlab{Fourier}
\end{equation}
In a homogeneous medium with refraction index $n$ the vector potential can be calculated~\cite{Jackson} from
\begin{align}
 [k^2-\frac{n^2\omega^2}{c^2}]\,\Phi(\mathbf{k},\omega)&=\frac{4\pi}{n^2}\rho(\mathbf{k},\omega)
  \nonumber \\
 [k^2-\frac{n^2\omega^2}{c^2} ]\,\mathbf{A}(\mathbf{k},\omega)&=\frac{4\pi}{c}\mathbf{J}(\mathbf{k},\omega)
 \eqlab{wavevector} \; .
\end{align}
For the problem under consideration the charge density is
\begin{equation}
 \rho(\mathbf{r},t) = q(z)\, \delta(\mathbf{r}-\mathbf{v}t-\mathbf{a}) \;, \eqlab{chdens}
\end{equation}
and the current density is
\begin{equation}
 \mathbf{J}(\mathbf{r},t)=\mathbf{v}\,\rho(\mathbf{r},t) \;, \eqlab{curdens}
\end{equation}
with $\mathbf{v}$ the velocity of the charge,
$\mathbf{a}=-a\hat{x}$ the distance under the surface, and $q(z)$ the charge
of the particle. The vector potential can now be written as
\begin{equation}
 \mathbf{A(k},\omega)=4\pi\frac{\mathbf{v}}{c}\frac{\rho(\mathbf{k},\omega)}{k^2-\frac{n^2 \omega^2}{c^2}} \;, \eqlab{Akw}
\end{equation}
with
\begin{equation}
 \rho(\mathbf{k},\omega)=\frac{Q}{\beta c} \frac{e^{i k_x a}\sin{(\frac{L}{2}(k_z-\frac{\omega}{\beta c})})}{2\pi^2
(k_z-\frac{\omega}{\beta c})} \;. \eqlab{rho}
\end{equation}
Note that because $\mathbf{v}$ is only non-zero in the $z$-direction, $A_z$ is the only non-vanishing component of $\mathbf{A}$.

The transmission coefficient for the vector potential can be derived from the field equations across the boundary. In general this is not easy, however, for the special case of interest here, this can be done by considering the electric field in the $y=0$ plane.
For electric fields generated by a vector potential of the form $A_i(r,t)=A_i^0 e^{i(k_xx+k_yy+k_zz-\omega t)}$ we obtain $ E_i=\partial_0 A_i - \partial_i A_0$, and thus $E_{||} = E_x\hat x \cdot \hat{||}+E_z\hat z \cdot \hat{||} =i\frac{\omega}{c}\frac{k_z}{k} A_x^0-i\frac{\omega}{c}\frac{k_x}{k}A_z^0$. For the field of the transmitted wave we may write $E_{||}' = t_{||} E_{||}$.
In our treatment it is sufficient to calculate $A_z'$ since only $A_z$ is non-vanishing,
\begin{equation}
 t_{A_z||} = \frac{{A_z^0}'}{A_z^0} = \frac{k_x}{k E_{||} } \frac{E'_{||} k'}{k_x'}  = t_{E||}   \frac{k_x}{k} \frac{k'}{k_x'} \;, \eqlab{deft}
\end{equation}
where $t_{E||}=2\cos \theta_i/(n'/n \cos \theta_i + \cos \theta_t)$ (see \figref{diagram} for the definition of the angles) is the transmission coefficient of the electric field. At this point we have made the implicit assumption that the observer is far from the surface and that the outgoing waves can be treated as plane waves.
We still perform a complete integral over all waves leading from the source to the surface.
For transmission parallel to the surface we thus derive that
\begin{equation}
 t_{A_z||}= \frac{2 k_x^2}{k_x k_x' + \frac{n^2}{n'^2}k_x'^2} \;, \eqlab{tAz}
\end{equation}
where $k_x^2=\frac{n^2\omega^2}{c^2}-k_y^2-k_z^2$ and $k_x'=k_x \sqrt{1-\frac{(n^2-n'^2)\omega^2}{c^2 k_x^2}}$.

An incoming wave $e^{i\mathbf{k \cdot r} - i\omega t}$ generates for $x>0$ a transmitted
wave $t_{A_z||} e^{i \mathbf{k' \cdot r}- i\omega t}$ where $t_{A_z||}$ is given by \eqref{tAz}.
The transmitted radiation ($x>0$) can now be expressed as
\begin{equation}
 {A}_z'(\mathbf{r},t) = \int \! \frac{d^3 k\,d\omega}{4\pi^2} \frac{2 k_x^2}{k_x k_x' + \frac{n^2}{n'^2}k_x'^2} A_z(\mathbf{k},\omega)  e^{i\mathbf{k'\cdot r}-i\omega t} \;, \eqlab{A3rt}
\end{equation}
where ${A}_z(\mathbf{k},\omega)$ is given by \eqref{Akw} and $\mathbf{k'}$ is a function of $\mathbf{k}$.

To evaluate this expression, first a change of variables is made to wave vectors in the $n'$ medium, $k_x' = \sqrt{k_x^2-\frac{n^2-n'^2}{c^2}\omega^2}$, $k_y'=k_y$, $k_z'=k_z$. Subsequently $\omega$ is integrated over by extending the integral into the complex plane where the contributions of the poles of $\frac{1}{k'^2-\frac{n'^2\omega^2}{c^2}}$ in $A_z(\mathbf{k},\omega)$ have to be considered. The poles are calculated by adding an infinitesimal imaginary part to account for causal propagation.
This leads to the limitation $\beta c t > L/2$. As the next step the $d^3 k'$ integral is written as $dk' \sin{\theta}\, d\theta \, d\phi$  where $k_x'=k'\cos\theta$, $k_y'=k'\sin\theta \sin \phi$ and $k_z'=k'\sin\theta \cos\phi$. The integrals can now be reduced to two integrals of the form $\int_{-\infty}^{\infty} dk'\, e^{i k' f(\theta,\phi)}$. After integrating over $k'$ this gives a delta function which can be used to perform the integral over $\phi$. The field can now be expressed as
\begin{equation}
 A_z(\mathbf{r},t)=A_z^{(1)}(\mathbf{r},t)+A_z^{(2)}(\mathbf{r},t)
\end{equation}
where the first contribution is
\begin{widetext}
\begin{equation}
 A_z^{(1)}(\mathbf{r},t)= \frac{-Q}{\pi n'} \int \sin{\theta} \,d\theta
 \frac{W(\theta) \,(z+\frac{L}{2})+\frac{\eta^2}{n' \beta} }
 {W^2(\theta) + \frac{2 W(\theta)\, (z+\frac{L}{2})}{n'\beta}+\frac{\eta^2}{n'^2\beta^2}-y^2\sin^2\theta}
 T_4 \, Re\left[ \frac{1}{\sqrt{\eta^2 \sin^2\theta - W^2(\theta)} } \right] \;,
\end{equation}
and the second contribution is
\begin{equation}
 A_z^{(2)}(\mathbf{r},t)= \frac{Q}{\pi  n'} \int \sin{\theta}\, d\theta
 \frac{\tilde{W}(\theta) \,(z-\frac{L}{2})+\frac{\tilde{\eta}^2}{n'\beta}}
 {\tilde{W}^2(\theta) + \frac{2 \tilde{W}(\theta) \, (z-\frac{L}{2})}{n'\beta} +\frac{\tilde{\eta}^2}{n'^2\beta^2}-y^2\sin^2\theta}
T_4\, Re\left[ \frac{1}{\sqrt{\tilde{\eta}^2 \sin^2\theta - \tilde{W}^2(\theta)} } \right] \;, \eqlab{A2rt}
\end{equation}
\end{widetext}
where
\begin{align}
 W(\theta)&=x \cos\theta  + a \cos\theta \sqrt{1+\frac{n^2-n'^2}{n'^2 \cos^2\theta}} -  ct/n' -\frac{L}{2 n' \beta} \;,
 \nonumber \\
 \tilde{W}(\theta)&=x \cos\theta  + a \cos\theta \sqrt{1+\frac{n^2-n'^2}{n'^2 \cos^2\theta}} -  ct/n' +\frac{L}{2 n' \beta} \;,
 \nonumber \\
 T_4&=\frac{2 \cos \theta}{\cos \theta \sqrt{1+\frac{n^2-n'^2}{np^2 \cos^2\theta}}+\frac{n^2}{n'^2}\cos \theta} \;,
 \nonumber \\
 \eta^2&=(z+\frac{L}{2})^2+ y^2 \;,
 \nonumber \\
 \tilde{\eta}^2&=(z-\frac{L}{2})^2+ y^2 \;,
\end{align}
Note that only the real part of the square root contributes because of the restriction $W(\theta)\leq
\eta \sin{\theta}$ which is imposed by the $\delta$-function in the derivation.

\subsection{Checking limiting cases}

\begin{figure}
 \centerline{ \includegraphics[width=0.47\textwidth]{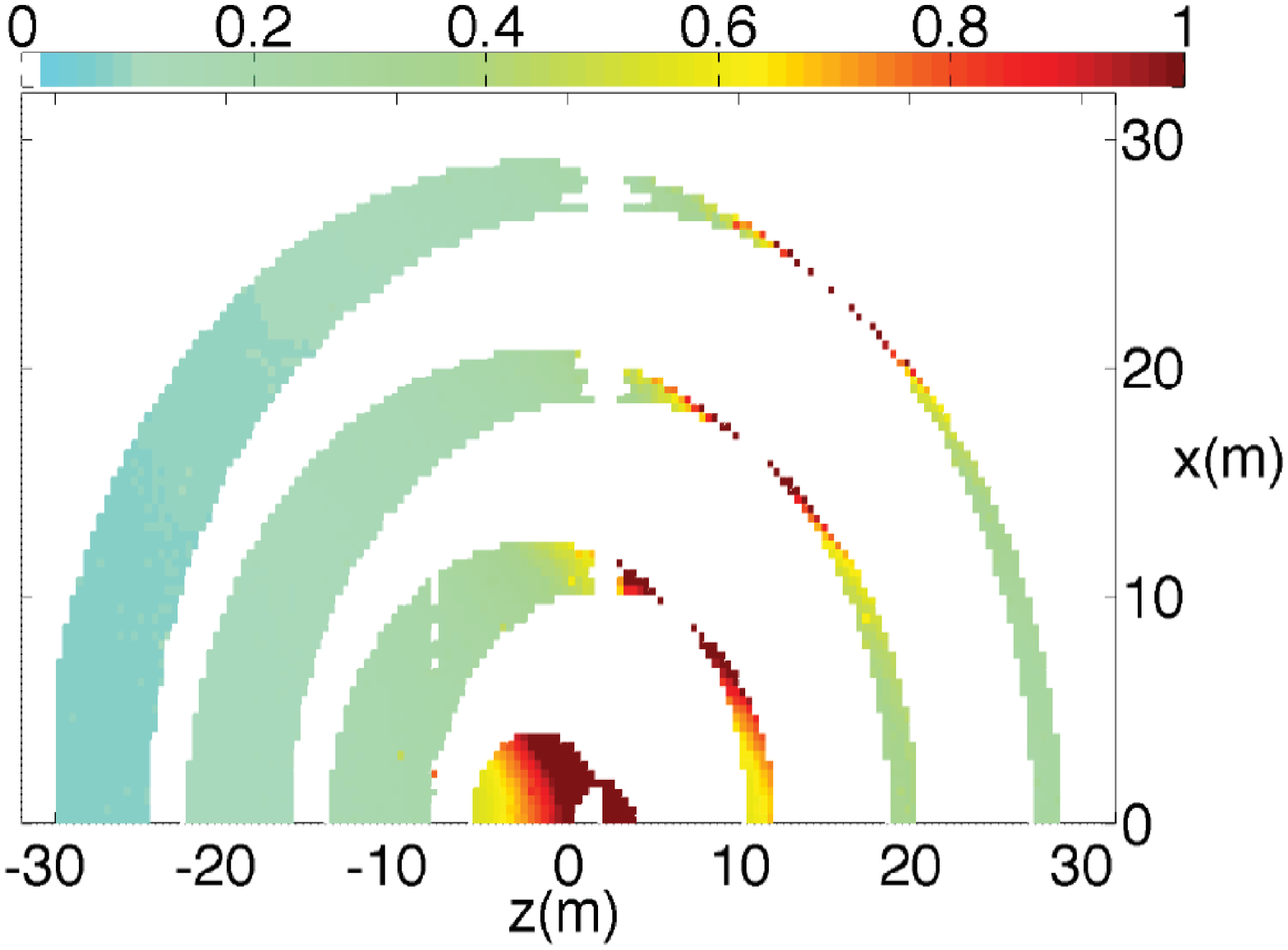} }
\caption{[color online] The vector potential $A_z$ for $y=0$ in the near field at
different times $ct$=5, 9, 13, 17\,m; $n=n'$=1.8, $a$=0, and $L$=4 as a function of $x$
and $z$. The absence of field near $z=2$ is due to numerical difficulties.} \figlab{growingfield}
\end{figure}

In \figref{growingfield} the vector potential is plotted in the $y=0$ plane for different times shortly after creation for a
homogeneous medium. It shows an outgoing pulse traveling at the light velocity  which is stronger and of shorter duration in the direction of the \v{C}erenkov cone. At angles away from the \v{C}erenkov angle the waves emitted from different parts for the current distribution no longer arrive at the same time. This is reflected in \figref{growingfield} by a broadening of the pulse which has the immediate consequence that the signal is coherent only for lower frequencies.

\begin{figure}
 \centerline{ \includegraphics[width=0.47\textwidth]{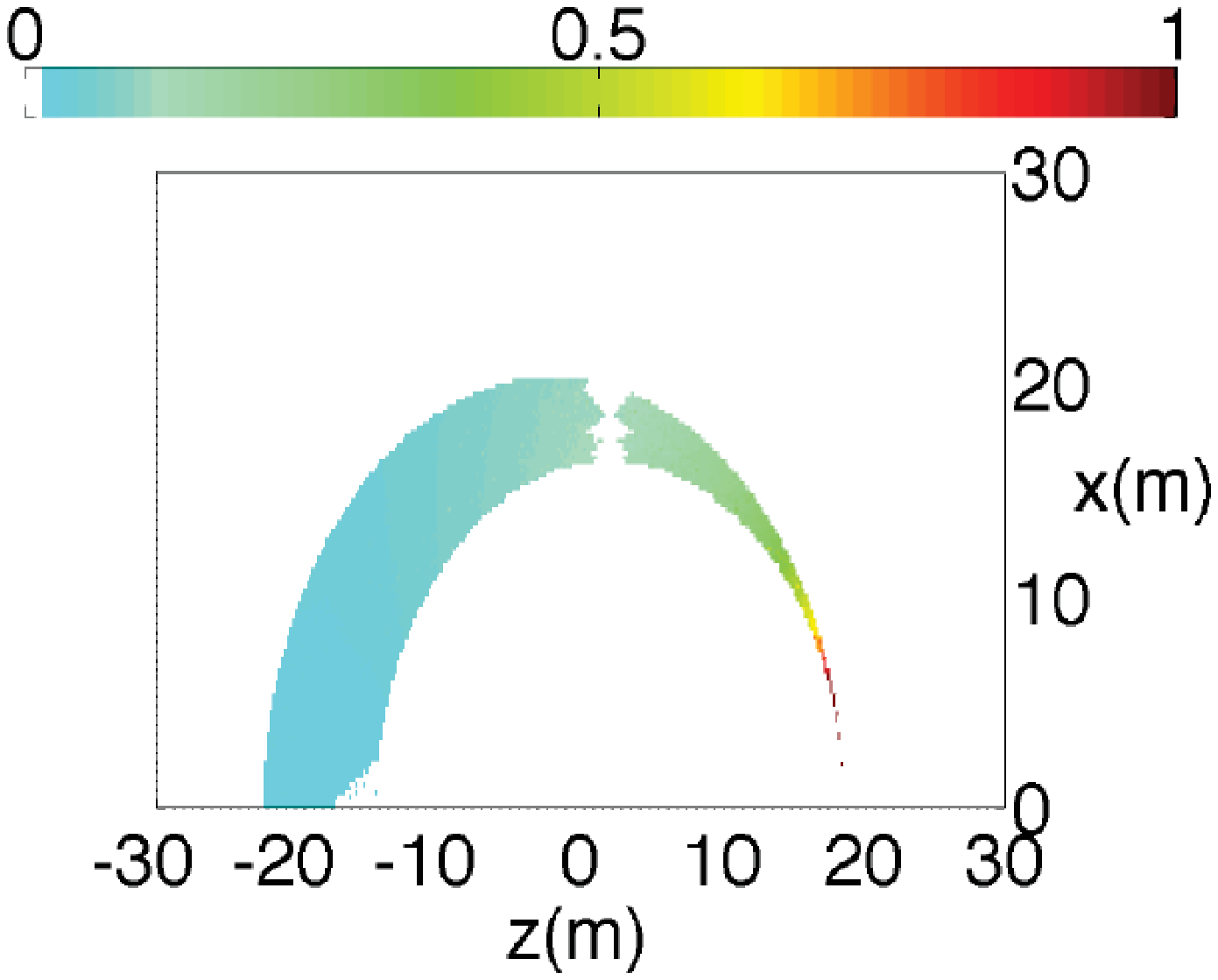} }
\caption{[color online] $A_z$ for $n$=1.8, $n'$=1.0 at $ct$=20\,m, $L$=4, and $a$=1. The absence of field near $z=2$ is due to numerical difficulties.}
\figlab{field10}
\end{figure}

When we include refraction into vacuum the outgoing wave has a different structure as shown in \figref{field10}. Because of the refraction at the surface the \v{C}erenkov angle is projected at zero degrees, however, there is still radiation transmitted through the surface.

The electric field can be calculated from the vector potential using $E(\mathbf{R},\omega) =\omega \sin{\theta} (A_z^{(1)}(\mathbf{R},\omega) + A_z^{(2)}(\mathbf{R},\omega))$. The Fourier transform is calculated numerically by taking the sum $A_3^{(i)}(\mathbf{R},\omega)=\sum_{t=t_0}^{t_1} e^{i\omega t}A_3^{(i)}(\mathbf{R},t)\Delta t/\sqrt{2\pi}$ where $A_3^{(i)}(\mathbf{R},t)$ is calculated numerically at each point. For a shower far below the surface ($a>>L$) viewed from far away ($R>>a$) this should give the same result as the analytic result \eqref{method}. In \figref{calccheck} both the double far field analytic result and the exact numerical result are compared for two different observing frequencies, showing that both results are practically indistinguishable.

\begin{figure}
\begin{center}
\includegraphics[width=.95\linewidth]{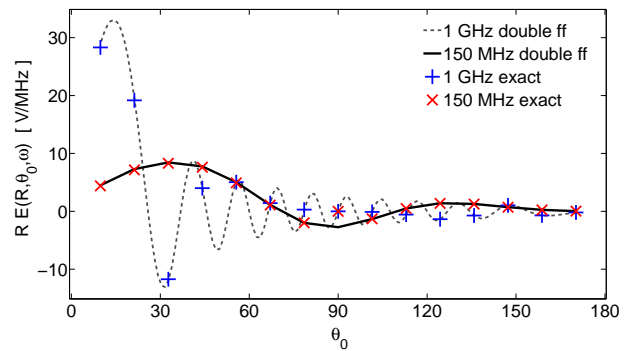}
\end{center}
\caption{[color online] Electric field strength as a function of angle at 150\,MHz and 1\,GHz at a distance of 1400\,km for a shower length of 3\,m, and a charge of $Q=10^{12}$e, calculated according to the double far-field analytic (see \secref{far}) and the exact numerical (see \secref{exact}) methods.} \figlab{calccheck}
\end{figure}

\subsection{Formation zone effects}

After having checked the calculation in the far-field regime we can use it to calculate the shower at distances close to the surface. The result of the calculation is shown in \figref{Efield_depth} for 3 different angles. This shows that within an accuracy of 1.5\% the field is the same for showers at depths ranging from 1\,cm to 1\,km when observed from sufficiently far away. The error we attribute to the numerical calculation and as a second order effect in $a/R$ and is largest for the calculation for $a=1$\,km. The important implication of this is that for an observer at Earth the observed electric field is independent of depth below the lunar surface. There is no shallow shower effect which supports the conclusion arrived at in \secref{zero}.

\begin{figure}
\begin{center}
\includegraphics[width=.85\linewidth]{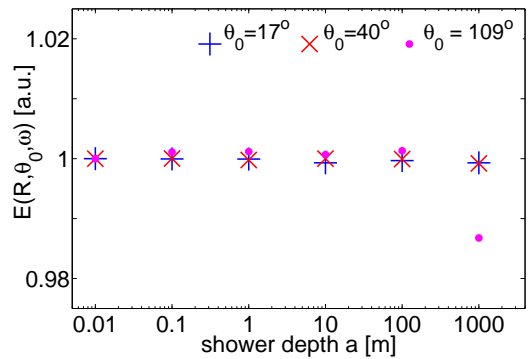}
\end{center}
\caption{[color online] Electric field strength as a function of depth, normalized to that at $a=10^{-2}$\,m, for 3 different angles at 150\,MHz at a distance of 1400\,km for a shower length of 3\,m.} \figlab{Efield_depth}
\end{figure}

\section{New limit on the UHECR flux}

As a first application we will use the present result to obtain limits on the flux of UHE cosmic rays using the results presented in a recent publication of the `NuMoon' observations of the Moon using the Westerbork Synthesis Radio Telescope (WSRT)~\cite{Sch09}.
The WSRT consists of an array of 14 parabolic antennas of 25~m diameter on a
2.7~km East-West line. In the observations we used the Low Frequency Front Ends
(LFFEs) which cover the frequency range 115--180~MHz with full polarization
sensitivity. The Pulsar Machine II (PuMa II) backend~\cite{kss08} can record a
maximum bandwidth of 160~MHz, sampled as 8 subbands of 20~MHz each. Only 11 of
the 12 WSRT dishes with equal spacing were used for this experiment.
In these observations as part of the NuMoon project, the radio spectrum was searched for short, nano-second pulses emitted from showers induced in the lunar regolith by UHE neutrinos. The data allowed a tightening of the bounds on the neutrino flux at high energies~\cite{Sch09}.

When an UHE neutrino interacts, most of the energy is carried away by the emerging lepton which, in general, does not produce a detectable signal while only about 20\% of the energy is deposited in a hadronic shower which emits a signal that can be detected at Earth.
When a cosmic ray impinges on the lunar surface all its energy will be converted into a hadronic cascade of energetic particles. This cascade will commence right at the surface and the shower maximum is thus within meters from the surface. Due to the absence of a formation zone these events should thus also give a signal of the characteristics of the ones that were searched for in the observations of Ref.~\cite{Sch09}. With its surface area of the order of $10^7$\,km$^2$, our finding of the absence of a formation zone thus shows that the Moon can be used as a sensitive cosmic-ray detector.

As the first step in the WSRT observations the narrow band Radio Frequency Interference (RFI) is filtered from the recorded time series data for each subband (with a sampling frequency of 40 MHz) and the dispersion due to the ionosphere of the Earth is corrected for.
Short, nano-second, pulses emitted from the Moon correspond to strong pulses with large bandwidth. To search for these pulses 5-time-sample-summed power spectra (so called $P_5$-spectra) were constructed for all subbands.
The data were kept for further processing when in all four subbands beamed at the same side of the Moon a value for $P_5$ larger than a certain threshold was found within a certain maximum time offset.
In a subsequent analysis additional constraints were imposed, such as eliminating pulses that had a long time duration and pulses that were found in both beams within a certain time limit.
The largest remaining pulse had a strength of $S=152$\,kJy which is a factor 3 larger than
expected for pure statistical noise.
To account for dead-time issues and filtering inefficiencies a complete simulation of the data taking was performed where short pulses with a random time offset have been added to the raw data (considered
as background). The detection efficiency (DE) was determined as the fraction of inserted pulses
that is retrieved after applying all trigger conditions and cuts that were used in the analysis.
The System Equivalent Flux Density (SEFD) for the WSRT, averaged over the frequency range under consideration, is $\sigma^2=400$~Jy per time sample.

A more detailed description of the followed procedure can be found in Ref.~\cite{Sch09} where it is concluded that in  46.7\,hours of observation no pulses from the Moon were detected with a strength exceeding $240$\,kJy with a 87.5\% probability.
Simulation calculations have been performed to convert this to a flux limit. In the simulation cosmic rays of a certain energy $E_s$ hit the lunar surface at arbitrary angles and create a particle cascades.
Based on Monte Carlo simulations the intensity of radio-waves emitted from such a cascade in the lunar regolith have been parameterized as function of emission angle and frequency~\cite{Zas92,Alv01,Gor04}. The emitted radiation is passed through the lunar surface following the usual laws of wave refraction which, due to the absence of formation zone effects, also applies to the case where the particle cascade occurs just below the surface. From the surface of the Moon to Earth the intensity follows the usual inverse square law. The details of the simulations are described in detail in Ref.~\cite{Sch06}. Using the model-independent procedure described in Ref.~\cite{Leh04} the non-observation of pulses of a certain strength can thus be converted in an energy-dependent 90\% confidence limit on the cosmic-ray flux as shown in \figref{wsrtlimits} by running the simulation for different values of $E_s$. In the simulations the effects of surface roughness can be ignored~\cite{Sch06}.

\begin{figure}
\centering
\includegraphics[width=.7\linewidth,viewport=37 140 510 660,clip]{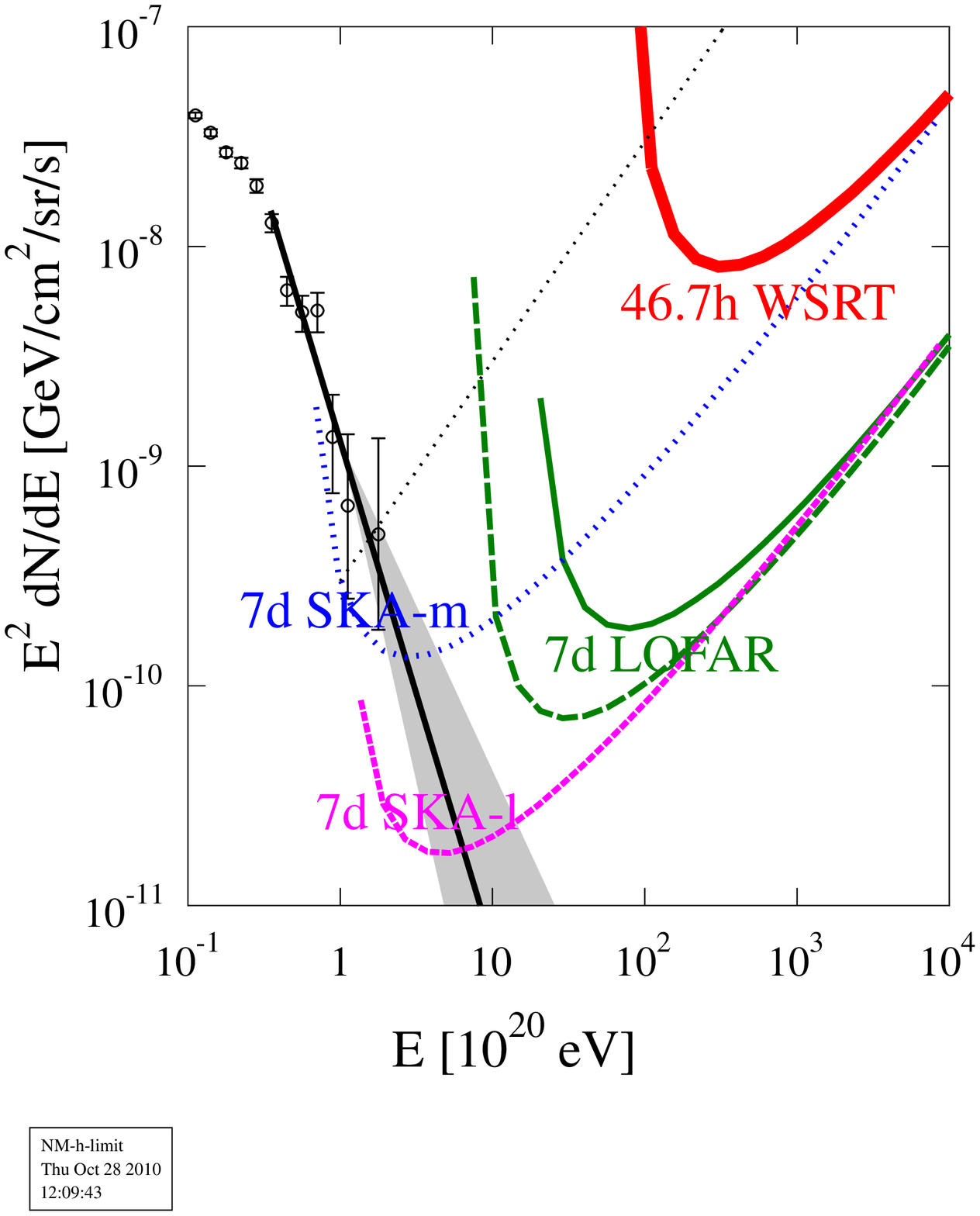}
\caption{[color online] The currently established cosmic-ray flux limit from WSRT observations~\cite{Sch09} (thick red drawn line) is compared to
the flux determined by the Pierre Auger Observatory~\cite{Abraham:2010mj} (data points with error bars) and a simple polynomial expansion (black line, see text). Also the prospective flux sensitivities are indicated that can be obtained with LOFAR~\cite{LOFAR} and SKA~\cite{SKA} observations.}
\figlab{wsrtlimits}
\end{figure}

The present limit for the flux of cosmic rays that follows from the existing WSRT observations~\cite{Sch09} is well above what could be expected based on the observations made by the Pierre Auger Observatory~\cite{Abraham:2010mj}. The thin dotted straight line in \figref{wsrtlimits} shows the model-independent differential flux limit, comparable to the presently set limits, for observations of the Pierre Auger Observatory, showing that the limit set by the WSRT observations is considerably lower albeit at considerably higher energies. The thick line corresponds to a polynomial expansion $~E^{-4.3}$ as advocated in Ref.~\cite{Abraham:2010mj} where the grey band corresponds to an uncertainty in the exponent of 0.8. The lower limit of the exponent has been taken from Ref.~\cite{HiRes}. The quoted values for the flux in Ref.~\cite{HiRes} lie well above the ones shown in \figref{wsrtlimits} which might be due to an uncertainty in the energy calibration~\cite{Abraham:2010mj}.
Future observations with new-generation radio telescopes such as LOFAR~\cite{LOFAR} or SKA~\cite{SKA} should reach much higher sensitivity for pulse detection, resulting in correspondingly lower energy thresholds as shown in \figref{wsrtlimits}. We show the sensitivity that can be reached in a one week measurement using the LOFAR telescope where the drawn curve uses only the core stations (SEFD of 93\,Jy, using 50\,MHz bandwidth) while the long-dashed curve uses all E-LOFAR stations (SEFD of 30\,Jy, using full bandwidth). We have assumed here a 100\% moon coverage and a detection threshold of $6\sigma$ where $\sigma$ is the amplitude of the noise.
In a one week measurement with the future SKA telescope (SEFD of 1.8\,Jy) the results depend on the frequencies used for the observation.
At lower frequencies (100-300\,MHz band, SKA-l in \figref{wsrtlimits}) one is sensitive to a smaller flux while at intermediate frequencies (300-500\,MHz band, SKA-m in \figref{wsrtlimits}) one is sensitive to cosmic-rays of lower energy. The increased sensitivity will make this method sensitive to cosmic ray energies of the order of $10^{20}$\,eV where, due to the large collecting area, competitive measurements are possible.

\section{Conclusions}

We have shown that the concept of a formation zone does not apply for the emission of electromagnetic radiation from a moving charge distribution over a finite distance inside a dielectric emitting \v{C}erenkov radiation. In particular we have considered the system where the charges move at close proximity to the surface separating the dielectric and vacuum. We have shown that the radiation penetrating the surface is independent of the distance of the charge distribution to the surface even for distances that are much smaller than the wavelength, provided the observer is sufficiently far away from the surface. In principle we have shown the absence of a formation zone only for a charge distribution with a block profile, however, the superposition principle can be used to show that this conclusion extents to showers with a realistic profile.

One field where this finding has a large impact is in the calculation of the acceptance of large scale cosmic-ray and neutrino detectors. As one application we have calculated the acceptance of the observations for the NuMoon project to cosmic rays and used this to derive a limit on the flux of cosmic rays for energies in excess of $2 \times 10^{22}$\,eV. If there would have been a formation length of the order of the wavelength of the observed radiation, the acceptance would be vanishingly small. We instead find that this approach will offer a competitive means of detecting the flux of cosmic rays with energies in excess of $10^{20}$\,eV with future synthesis radio telescopes.

\begin{acknowledgments}
This work was performed as part of the research programs of the Stichting voor Fundamenteel Onderzoek der Materie (FOM), with financial support from the Nederlandse Organisatie voor Wetenschappelijk Onderzoek (NWO), and an advanced grant (Falcke) of the European Research Council.
\end{acknowledgments}

\end{document}